\documentclass[preprint,tightenline,aps,amsmath,amssymb]{revtex4}
\usepackage{dcolumn}% Align table columns on decimal point
\usepackage{bm}% bold math
\usepackage[dvips]{graphicx}
\usepackage{wrapfig,subfigure}
\usepackage{amssymb}
\usepackage{color}

\begin{document}

%\twocolumn[

\title{Distribution of Fluctuational Paths in
Noise-Driven Systems\footnote{This paper is dedicated to
Rolf Landauer, with deep respect and admiration}}
\author{M.I. Dykman and V.N. Smelyanskiy}
\affiliation{
Department of Physics and Astronomy,
Michigan State University, East Lansing, MI 48824, USA}
%\date{\today}

\begin{abstract}
Dynamics of a system that performs a large fluctuation to a given state
is essentially deterministic: the distribution of fluctuational paths
peaks sharply at a certain {\it optimal} path along which the system is
most likely to move. For the general case of a system driven by colored
Gaussian noise, we provide a formulation of the variational problem for
optimal paths. We also consider the prehistory problem, which makes it
possible to analyze the shape of the distribution of fluctuational
paths that arrive at a given state. We obtain, and solve in the
limiting case, a set of linear equations for the characteristic width of
this distribution.
\end{abstract}
\maketitle

%\narrowtext
\section{Introduction}
Large fluctuations, although infrequent, play a fundamental role in a
broad range of processes, from diffusion in crystals to nucleation at
phase transitions, mutations in DNA sequences, and failures of
electronic devices. In many cases the fluctuating systems of interest
are far from thermal equilibrium.  Examples include lasers, pattern
forming systems
\cite{Hohenberg}, trapped electrons which display bistability and
switching in a strong periodic field \cite{Dehmelt1,Gabrielse}, and
spatially periodic systems (ratchets) which display a unidirectional
current when driven away from thermal equilibrium \cite{ratchets}.

It was very clearly shown by Landauer
\cite{Landauer,blowtorch} that,
whereas for systems in thermal equilibrium the probabilities of
fluctuations are known at least in principle, for nonequilibrium
systems there is no universal relation from which these probabilities
can be obtained: even though the system mostly stays in the vicinity
of one of locally stable states, the distribution over these stable
states can be found only from global analysis. This distribution may
be strongly affected by nonthermal perturbations in the rarely
occupied intermediate states, i.e. by the large fluctuations which
determine the probabilities of switching between the stable states
(Landauer's blowtorch theorem).

The major physical problems in the theory of large fluctuations are
not only calculation of the fluctuation probabilities, but also
analysis of the {\it dynamics} of large fluctuations. Understanding
this dynamics is particularly important for controlling large
fluctuations.

An intuitive approach to the theory of large fluctuations makes use of
the {\it optimal path} concept. The optimal path is the path along
which the system is most likely to move when it fluctuates to a given
state from the vicinity of the stable state. Although trajectories
of a fluctuating system are random, it is clear from Fig.~1 that the
probabilities for the system which is found at a point $q_f$ at an
instant $t_f$ to have arrived to this state along different paths are
very different: e.g., it is unlikely that the system has been staying
far from the equilibrium position for a long time, or that it has
experienced an extremely large acceleration, as is the case for the
paths 1 and 3 in Fig.~1.

Optimal paths and the probability distribution of the fluctuational
paths are real physical objects: they have been
observed in analog experiments
\cite{prehistory,corrals,nature} and digital simulations \cite{Morillo}.
In the theory of large fluctuations, the pattern of optimal paths
plays a role similar to that of the phase portrait in nonlinear
dynamics.

The fundamental role of the distribution of fluctuational paths was
recognized already by Onsager and Machlup \cite{Onsager}; in fact,
they obtained optimal paths for a linear system in thermal equilibrium
with the bath with a short correlation time (the approximation of
Brownian motion). For such systems, whether they are linear or
nonlinear, the optimal path to a given state is the time-reversed path
from this state to the vicinity of the stable state in the neglect of
fluctuations (the deterministic path)
\cite{nature,Marder}. This is no longer true for nonequilibrium systems,
because, in general, they lack time reversibility. Even for simple
nonequilibrium systems the pattern of optimal paths may have singular
features \cite{Jauslin,DMS}.
\begin{figure}[h]
\includegraphics[width=3.5in]{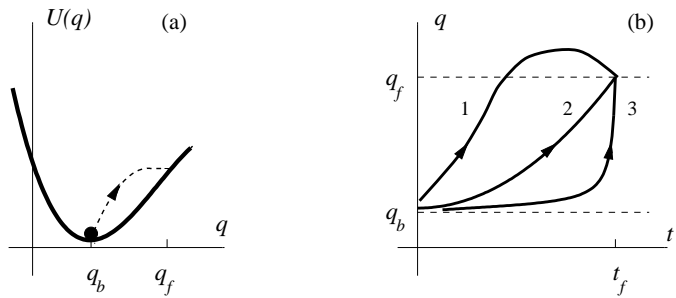}
\caption{(a) A particle with a coordinate $q$ fluctuating away from
the bottom $q = q_b$ of the potential well $U(q)$ to a point
$q_f$. (b) Various paths along which a particle can reach $q_f$ at a
given instant $t_f$ in the course of the fluctuation; the probability
densities for moving along different paths are exponentially
different. With overwhelming probability the system moves to a point
$(q_f,t_f)$ along an {\it optimal path}.}
\end{figure}

In the present paper we provide a general formulation of the problem
of optimal paths for nonequilibrium systems driven by Gaussian
noise. In Sec.~II, based on a path-integral expression for the
transition probability density, which allows for a prefactor, we
derive an integral variational functional for optimal fluctuational
paths. In Sec.~III we provide a formulation of the prehistory problem
for the distribution of fluctuational paths to a given state. This
formulation is reduced to a linear integro-differential equation; the
explicit form of this equation depends on the form of the correlation
function of the noise. In Sec.~IV we discuss the solution of this
equation in the case where the destination state is close to the stable
state of the system. Sec.~V contains concluding remarks.

\section{Large fluctuations induced by Gaussian
noise with an arbitrary power spectrum}

Nonlinear systems driven by nonthermal Gaussian noise form an important
and fairly general class of nonequilibrium systems. In spite of seeming
simplicity, they display a variety of interesting effects,
unidirectional current in periodic structures being an example
\cite{ratchets}.  Substantial progress in the theory of large
fluctuations in such systems has been made within the last two decades,
particularly by implementing the path integral technique  (see
\cite{DK79,Luciani,Bray,Pesquera,Dykman90} and references therein). So
far this technique has been applied to systems driven by the noise which
is a component of a Markov process \cite{Dykman90}, in which case the
reciprocal power spectrum of the noise $1/\Phi(\omega)$ is a polynomial
in $\omega^2$, where
\begin{equation}
\Phi(\omega) = \int dt\,e^{i\omega t}\phi(t),\quad \phi(t)=
\langle f(t)f(0) \rangle. \label{power_spectrum}
\end{equation}
\noindent It is advantageous to express the characteristics of
fluctuations in the system in terms of the noise power spectrum, since
$\Phi(\omega)$ can often be measured in experiment. The shape of
$\Phi(\omega)$ depends on the source of the noise and the coupling
between the dynamical system and this source.

The noise with $\Phi^{-1}(\omega)$ of the form of a polynomial in
$\omega^2$, although interesting, is not the most general type of noise.
An important example of Gaussian noise with a non-polynomial
$\Phi^{-1}(\omega)$ is the noise with Gaussian power spectrum in the
central part
\begin{equation}
\Phi(\omega) = D\exp\left[ -\left(\omega^2 -
\Omega_0^2\right)^2/4\Omega_0^2\sigma^2\right],\;
|\omega^2-\Omega_0^2| \alt \sigma^2.
\label{gaussspect}
\end{equation}
\noindent
In particular, noise of this
sort is produced by the electric field of  inhomogeneously
broadened (in particular, Doppler-broadened) radiation.

\subsection{Transition probability in a system driven by Gaussian noise}

In this section we provide a general formulation of the problem of
large fluctuations induced by Gaussian noise. We consider stationary
systems or systems in a time-periodic field. To simplify notations, we
will assume that the system under consideration is described by one
dynamical variable, $q$. The Langevin equation of motion is then of
the form:
\begin{equation}
\dot q = K(q;t) + f(t),\quad K(q;t+T) = K(q;t),\label{Langevin}
\end{equation}
\noindent
where $f(t)$ is the zero-mean stationary noise.
We assume that the noise is characterized by a certain correlation
time $t_{corr}$ over which its correlation function decays
(at least exponentially, in the limit of large time).

For weak noise intensities, over a time $t$ which exceeds $t_{corr}$
and the characteristic relaxation time in the absence of noise
$t_{rel}$, the system will approach the stable state $q^{(0)}(t)$ and
will then perform small fluctuations about it. In periodically driven systems
the state $q^{(0)}(t)$ is also periodic and is given by the equation
\begin{equation}
\dot q^{(0)} = K(q^{(0)};t),\qquad q^{(0)}\left(t+T\right)
= q^{(0)}(t).
\label{attractor}
\end{equation}
\noindent
(we assume that the period of the state $q^{(0)}$ is the same as that of
the periodic driving).

In the course of a large fluctuation, the dynamical system is brought
from the attractor to a distant point $q_f$ at the instant $t_f$
(cf. Fig.~1). For this to happen the system should have been subjected
to {\em finite} forcing over certain time. Different
realizations of the force $f(t)$ can result in the same final state.
The system trajectories $q(t)$ for each realization of $f(t)$ are
deterministic rather than random, and they are independent of the
characteristic noise intensity
\begin{equation}
D = {\rm max}\;\Phi(\omega).\label{D_defined}
\end{equation}

The probability density of realizations of $f(t)$ is given
by the functional (cf. \cite{Feynman})
\begin{equation}
{\cal P}[{f}(t)] = \exp\left[-{1\over 2D}\int dt\,dt'\,
f(t)\hat{\cal F}(t-t')f(t')\right],\label{prob_fnctnl}
\end{equation}
\noindent
where the operator $\hat {\cal F}(t)$ is related to the correlation
function of the noise $\phi(t)$ by the expression
\begin{equation}
\int dt_1\,\hat {\cal F}(t-t_1)\phi(t_1-t') = D\delta(t-t').
\label{F_vs_phi}
\end{equation}
\noindent
In some cases (in particular, for the noise $f(t)$ being a component
of a Markov process) a formal solution of this equation can be written
as
\begin{equation}
\hat{\cal F}(t) = D\delta(t)/\Phi(-id/dt).\label{formal}
\end{equation}
\noindent
Here, we have taken into account that the noise correlation function is
even, $\phi(t) = \phi(-t)$, as is also the noise power spectrum,
$\Phi(\omega) = \Phi(-\omega)$.

One can write the probability density $p(q_f,t_f)$ for the noise-driven
system to arrive at the point $q_f$ at the instant $t_f$, provided it
has started from the point $q_i$ at the initial instant $t_i$,
as a path integral
\begin{equation}
p(q_f,t_f) = \left\langle\int_{q_i
\approx q^{(0)}(t_i)}^{q_f}{\cal D}q(t)\,
\delta\left[q(t)-q_{det}(t;f|q_i)\right]\right\rangle. \label{density_general}
\end{equation}
\noindent Here, $q_{det}(t;f|q_i)$ is the solution of the dynamical
equation of motion (\ref{Langevin}) for a given realization of the noise
$f(t)$, and $\delta[q(t)-q_{det}(t)]$ is the functional delta-function:
it peaks at the function $q(t)$ equal to $q_{det}(t)$. The averaging
$\langle\ldots\rangle$ means integration over $f(t)$ with the
probability density functional (\ref{prob_fnctnl}) as a weighting factor
\cite{Feynman}. In what follows we assume that $t_i\rightarrow -\infty$
and the initial point is close to the attractor, $q_i \approx
q^{(0)}(t_i)$. In this case the function $p$ (\ref{density_general})
gives the stationary probability distribution, which is periodic in time
for the time-periodic force $K$ in (\ref{Langevin}), $p(q_f,t_f) =
p(q_f,t_f+T)$.

It is convenient to perform averaging over $f$ in
(\ref{density_general}) by writing the $\delta$-function in the form of
a path integral over an auxiliary variable $k(t)/D$. Using standard
transformations \cite{Feynman,Phythian,Luciani} one can show that the
expression for $p(q_f,t_f)$ can be  written in the following form:
\begin{eqnarray}
p(q_f,t_f) &=& C\int {\cal D}f(t)\,{\cal P}[f(t)]\int {\cal D}{k(t)\over D}
{\cal D}q(t)\,\label{complete_p}\\
&\times &\exp\left\{\int_{t_i}^{t_f} dt\left[i{k(t)\over D}
\left[\dot q - K(q;t) - f(t)\right]
-{1\over 2}K'\right]\right\},\quad
K'(q;t)\equiv {\partial K(q;t)\over \partial q},
\nonumber
\end{eqnarray}
\noindent where $C$ is the normalization constant, and ${\cal P}[f(t)]$
is the probability density functional for the random force
(\ref{prob_fnctnl}).

\subsection{Variational problem for optimal paths}

If $D$ is sufficiently small, as we assume, then, for all $ f(t)$ which
result in a large fluctuation to a given state, the values of the
probability density functional (\ref{prob_fnctnl}) are exponentially
small; they also exponentially strongly differ from each other for
different $f(t)$.  Thus one would expect that there exists one realization
$f(t)=f_{opt}(t)$ which is exponentially more probable than the others.
This realization provides the maximum to the functional
(\ref{prob_fnctnl}) subject to the {\it constraint} that the system
described by Eq.~(\ref{Langevin}) is driven to a designated state $q_f$.
Respectively, there are in fact {\bf two} interrelated through
({\ref{Langevin}) optimal paths: that of the system, $q_{opt}(t)$, and
that of the force, $f_{opt}(t)$ \cite{DK79,Dykman90}.

Formally, optimal paths can be obtained for small $D$ by
evaluating the path integral (\ref{complete_p}) by the steepest
descent method. It follows from Eqs.~(\ref{prob_fnctnl}),
(\ref{complete_p}) that the optimal paths provide the minimum to the
functional
\begin{equation}
{\cal R}[q(t),\lambda(t),f(t)] = {1\over
2}\int\!\!\!\int_{-\infty}^{\infty} dt\,dt'\, f(t)\hat {\cal
F}(t-t')f(t') + \int_{t_i}^{t_f} dt\, \lambda(t)
\left[\dot q - K(q;t)-f(t)\right],
\label{varfunct}
\end{equation}
\noindent where $\lambda(t)\equiv -ik(t)$; one can think of
$\lambda(t)$ as of a Lagrange multiplier that relates to each other the
optimal realization of the random force and the path of the system.

Variational equations for the trajectories that provide an
extremum to the functional (\ref{varfunct}) are of the form
\begin{equation}
\int dt'\hat {\cal F}(t-t')f(t') -\lambda(t) =0, \quad
\dot\lambda(t) + K'(q;t)\lambda(t) = 0,\quad \dot q(t) -K(q;t)
-f(t) = 0.\label{var_eqn}
\end{equation}
\noindent In the problem of the stationary probability density for the
system  to be in the state $q_f$  at the time $t_f$ (this probability
density is periodic in $t_f$ with the period $T$), the boundary
conditions for Eqs.(\ref{var_eqn}) take the form (cf.~\cite{Dykman90})
\begin{eqnarray}
f(t) \rightarrow 0 \quad {\rm for} \quad t\rightarrow
\pm \infty,&&\quad \lambda(t) \rightarrow 0 \quad {\rm for} \quad
t\rightarrow - \infty,\quad \lambda(t)= 0 \quad {\rm for} \quad t>
t_f,\nonumber\\
&&q(t) \rightarrow q^{(0)}(t) \quad {\rm for} \quad
t\rightarrow - \infty,\quad q(t_f) = q_f.  \label{boundary}
\end{eqnarray}
\noindent In deriving the boundary conditions (\ref{boundary}) for
$t\rightarrow -\infty$ we took into account that the system fluctuates
about the stable state for a long time before the large fluctuation
starts (cf. Fig.~1). Respectively, one may set in
(\ref{density_general}) $t_i\rightarrow -\infty,\, q(t_i)=q^{(0)}(t_i)$
(this is consistent with Eqs.~(\ref{var_eqn}). On the other hand, the
motion of the system {\bf after} it has reached the point $q_f$ is not
important for the large fluctuation. Therefore the constraint on $f(t)$
is lifted for $t>t_f$. Clearly, the force decays to zero for $t>t_f$.

The boundary conditions should be modified if one considers the
problem of escape from a metastable state. By generalizing the
arguments \cite{Dykman90} to the case of a non-polynomial reciprocal
power spectrum $1/\Phi(\omega)$, one can show that the optimal escape
path corresponds to the system approaching the unstable periodic state
$q_{\rm u}(t)$ (the saddle state) for $t_f\rightarrow \infty$, and
$\lambda(t_f)\rightarrow 0$ for $t_f\rightarrow \infty$, because in
this case $\int_0^T dt\,K'(q_{\rm u}(t);t) > 0$. The analysis of the
escape problem is beyond the scope of the present paper.

Eqs.~(\ref{var_eqn}), (\ref{boundary}) provide a complete set of
equations for the interconnected optimal fluctuational paths of the
system and the force, $q_{opt}(t|q_f,t_f)$ and $f_{opt}(t|q_f,t_f)$,
for reaching the state $q_f$ at the instant $t_f$. The probability to
reach this state, according to Eq.~(\ref{complete_p}), is of the form
\begin{equation}
p(q_f,t_f)\propto \exp\left[- R(q_f,t_f)/D\right],\; R(q_f,t_f) = {\cal
R}[q_{opt},\lambda_{opt},f_{opt}]\equiv {\rm min}\,{\cal
R}[q,\lambda,f].\label{W(q_f)}
\end{equation}
\noindent
where $\lambda_{opt}(t)$ is the optimal Lagrange multiplier as given
by Eqs.~(\ref{var_eqn}), (\ref{boundary}).

We note that Eqs.~(\ref{var_eqn}), (\ref{boundary}) may have several
solutions. In this case, the physically meaningful solution is the one
that provides the {\it absolute minimum} to the functional ${\cal R}$.
The criterion of applicability of the approach is $R/D\gg 1$ - it is
this condition that determines how small should the noise intensity
$D$ be.

An important feature of fluctuations induced by non-white noise, as it
is clear from (\ref{var_eqn}), (\ref{boundary}), is that the optimal
force $f_{opt}(t)$ does not become equal to zero once the system has
reached the state $q_f$. Time evolution of the optimal force is given by
the equation
\begin{equation} f_{opt}(t) =
\int_{-\infty}^{t_f}dt'\,\bar\phi(t-t')\lambda_{opt}(t'), \quad
\bar\phi(t) = D^{-1}\phi(t). \label{f_general}
\end{equation}
\noindent The function
$\bar\phi(t)$ is the noise correlation function rescaled so that it was
independent of the characteristic noise intensity $D$ (the noise
correlator $\phi \propto D$, cf. (\ref{power_spectrum}),
(\ref{D_defined})).

Based on Eq.~(\ref{f_general}) one can predict how the system will move,
most likely, {\it after} the state $q_f$ is reached. The trajectory of
the system is described just by the equation of motion (\ref{Langevin})
with the random force given by Eq.~(\ref{f_general}). We note that, even
for the state $q_f$ lying in the basin of attraction to the initially
occupied stable state from which the fluctuation starts, on the way back
from $q_f$ the system may come not necessarily to the same stable state,
but to a different state. This is in contrast with what happens in
systems driven by white noise, unless $q_f$ lies very close to the basin
boundary. An example of such fluctuations in systems driven by colored
noise was, in fact, considered in \cite{Dykman90},  although
trajectories of the system after it has fluctuated to a remote state were
not analyzed.

\subsection{Vicinity of the stable state}

To illustrate the solution of the variational problem (\ref{var_eqn}),
(\ref{boundary}) we will consider a simple case where the state $q_f$
is close to the stable state $q^{(0)}(t_f)$, so that the force
$K(q;t)$ can be linearized in $q-q^{(0)}(t)$, and yet the difference
$|q_f-q^{(0)}(t_f)|$ is big enough so that the asymptotic expression
(\ref{W(q_f)}) applies. With account taken of Eq.~(\ref{F_vs_phi}), we
obtain:
\begin{eqnarray}
&& \lambda(t) = u(t,t_f)\lambda_f,
\quad q(t)-q^{(0)}(t) = \int_{-\infty}^td\tau\,f(\tau)u(\tau,t),\quad
\label{lambda_q}\\
&&u(t,t')=\exp\left[\int_{t}^{t'} d\tau K'(q^{(0)}(\tau);\tau)\right].\nonumber
\end{eqnarray}

One can see from Eq.~(\ref{lambda_q}) that
\begin{equation}
\lambda_f= g^{-1}(t_f,t_f)\left[q_f-q^{(0)}(t_f)\right],\quad
g(t,t') = \int_{-\infty}^td\tau
\int_{-\infty}^{t'} d\tau'\,\bar\phi(\tau-\tau')u(\tau,t)
u(\tau',t').\label{G(t_f)}
\end{equation}
From Eqs.~(\ref{f_general}) - (\ref{G(t_f)}) one can find the
activation energy (\ref{W(q_f)}) of reaching the state $q_f$,

\begin{equation}
R(q_f,t_f) = {1\over 2}g^{-1}(t_f,t_f)\left[q_f-q^{(0)}(t_f)\right]^2.
\label{harmonic}
\end{equation}
\noindent The activation energy (\ref{harmonic}) is quadratic in the
distance between the state $q_f$ and the stable state $q^{(0)}(t_f)$.
The proportionality factor $g^{-1}(t_f,t_f)$ depends on the shape of the
noise power spectrum, and also on the dynamics of the system in the
absence of noise. We note that, if the regular force $K(q;t)$ is
independent of $t$, the stable state $q^{(0)}$ is
independent of $t$, and the function $g(t_f,t_f)$ is independent of $t_f$,
as expected.

\section{The prehistory problem}

The distribution of paths for large fluctuations can be investigated
and visualized through the analysis of the prehistory probability
density, $p_h(q_h,t_h|\,q_f,t_f)$ \cite{prehistory}. This is the
conditional probability density for a system that (i) had been
fluctuating about $q^{(0)}(t)$ for a time greatly exceeding the
relaxation time of the system $t_{rel}$ and the correlation time of
the noise $t_{corr}$, and (ii) arrived to the point $q_f$ at the
instant $t_f$, to have passed through (and been observed at) the point
$q_h$ at the instant $t_h$, $t_h < t_f$. Using the path-integral
formulation (\ref{complete_p}) one can write the prehistory
probability density as
\begin{eqnarray}
p_h(q_h,t_h|\,q_f,t_f)&=&
M\int {\cal D}f(t)\,{\cal P}[f(t)]\int {\cal D}{k(t)\over D}
\int_{q_i}^{q_f}{\cal D}q(t)\,\delta\left[q(t_h)-q_h\right]\label{pre_defined}\\
&\times &\exp\left\{\int_{t_i}^{t_f} dt\left[i{k(t)\over D}
\left[\dot q - K(q;t) - f(t)\right]
-{1\over 2}K'(q;t)\right]\right\},\nonumber
\end{eqnarray}
\begin{equation}
\int dq\, p_h(q,t|\,q_f,t_f) = 1.\label{normalization}
\end{equation}
\noindent  The normalization constant $M$ in the general expression for
$p_h$ (\ref{pre_defined}) is defined by the condition
(\ref{normalization}). Throughout this section we assume that
$t_i\rightarrow -\infty$.

We expect (and will confirm {\it a posteriori}) that, for small
$D$, the distribution $p_h(q,t|\,q_f,t_f)$ peaks sharply for $q$ lying
close to the optimal fluctuational path
$q_{opt}(t|q_f,t_f)$. Therefore in evaluating $p_h$ one can expand the
exponent in ${\cal P}[f]$ and the term with $k(t)$ in the exponent in
Eq.~(\ref{pre_defined}) in the deviations $\delta f(t), \delta q(t),
\delta k(t)$ from the optimal realizations of the random force $f_{opt}(t)$,
the trajectory $q_{opt}(t)$, and $k_{opt}(t)\equiv
i\lambda_{opt}(t)$.

In this paper we will not address the problem of singularities of
optimal paths in systems driven by colored noise \cite{Dykman90}, which
has been understood only recently for white-noise driven systems
\cite{DMS}. If the point $(q_f,t_f)$ is far from singularities, it
suffices to keep in the aforementioned expansion only the second-order
terms in $\delta f(t), \delta q(t), \delta k(t)$. Integrating over
$\delta f(t)$ and writing  $\delta[q(t_h)-q_h]$ as an integral, one can
re-write Eq.~(\ref{pre_defined}) in the following form
\begin{equation}
p_h(q_h,t_h|\,q_f,t_f)= M_1\int_{-\infty}^{\infty} d{a\over 2\pi
D}\int {\cal D}{k(t)\over D}
\int {\cal D}q(t)\,
\exp\left(-S[\delta k(t),\delta q(t)]/D\right), \label{p_h_transf}
\end{equation}
\noindent
where the quadratic functional $S$ is given by the expression
\begin{eqnarray}
&&S[\delta k,\delta q] = {1\over 2}\int\!\!\!\int_{-\infty}^{t_f}dt
dt'\,\delta k(t)\bar\phi(t-t')\delta k(t')\nonumber\\
&&\quad - i\int_{-\infty}^{t_f}dt\,
\delta k(t)[\delta \dot q(t) - K'(t)\delta q(t)] -
{1\over
2}\int_{-\infty}^{t_f}dt
\lambda_{opt}(t)K''(t)\delta q^2(t) +ia[q(t_h)-q_h],\label{quadratic}
\end{eqnarray}
\noindent
with
\begin{eqnarray}
&&\delta k(t)= k(t) - k_{opt}(t)\equiv k(t) -i\lambda_{opt}(t),\;
\delta q(t)= q(t) - q_{opt}(t),\nonumber\\
&& K'(t)\equiv K'(q_{opt}(t);t),\; K''(t)\equiv
K''(q_{opt}(t);t)= \partial^2 K/\partial q^2,\nonumber
\end{eqnarray}
\noindent
and with the boundary conditions
\begin{equation}
\delta q(-\infty)=\delta q(t_f) = 0,\quad \delta q(t_h)=
q_h -q_{opt}(t_h),\quad \delta k(-\infty)=0.
\label{b_c_prehistory}
\end{equation}

It is convenient to rewrite the expression for $S$ in the matrix form,
\begin{equation}
S[-i\psi_1,\psi_2]={1\over 2}\int\!\!\!\int_{-\infty}^{t_f}dtdt' \,
\mbox{\boldmath$\left(\right.$}\psi_1(t),\psi_2(t)
\mbox{\boldmath$\left.\right)$}
\hat {\cal H}(t,t')
\left( \begin{array}{c}\psi_1(t')\\ \psi_2(t') \end{array}\right)
+ ia[q(t_h)-q_h],
\label{matrix}
\end{equation}
\noindent
where $\hat {\cal H}$ is a Hermitian operator,
\begin{equation}
\hat {\cal H}(t,t') = \left(\begin{array}{ll}
-\bar\phi(t-t')& \delta(t-t')\left(K'(t')-d/ dt'\right)\\
\delta(t-t')\left(K'(t')+d/ dt'\right)\hspace*{5pt} &
-\delta(t-t')\lambda_{opt}(t)K''(t)
\end{array}\right).\label{kernel}
\end{equation}
\noindent
In obtaining Eq.~(\ref{kernel}) we took into account the boundary
conditions (\ref{b_c_prehistory}). As we will see, of immediate
interest is the value of $S[\delta k,\delta q]$ for purely imaginary
$\delta k$ and real $\delta q$, which justifies the unusual form of $S$
(\ref{matrix}).

\subsection{The variance of the prehistory probability distribution}

Integration over $\delta k(t),\delta q(t)$ in the expression
(\ref{p_h_transf}) can be performed by the steepest descent method. In
this method, one has to find the extremum of the quadratic functional
$S$, which requires solving the following equations for the extreme
values of $\delta k(t) = -i\psi_1(t), \delta q(t) = \psi_2(t)$:
\begin{equation}
\int_{-\infty}^{t_f}dt'\hat {\cal H}(t,t')
\left( \begin{array}{c}\psi_1(t')\\ \psi_2(t') \end{array}\right)
=- ia\delta(t_h-t)
\left( \begin{array}{c}0\\ 1 \end{array}\right),\quad
\psi_2(t_h) = q_h-q_{opt}(t_h)\label{matr_eqn}
\end{equation}
\noindent where the value of $a$ is determined by the boundary conditions
(\ref{b_c_prehistory}).

The solution of Eq.~(\ref{matr_eqn}) can be sought in terms of the Green
function $G_{ij}(t,t')$ which provides the solution to the equation
\begin{equation}
\int_{-\infty}^{t_f} dt'\,\hat{\cal H}_{ii_1}(t,t')G_{i_1j}(t',t'')=
\delta_{ij}\delta(t-t''), \quad G_{ij}(t,t') = G_{ji}(t',t),\quad
G_{i2}(t,t_f) = 0.\label{Green}
\end{equation}
\noindent
Here, summation is performed over repeated subscripts $i_1$; the
subscripts $i,i_1,j$ take on the values 1,2. The functions
$\psi_i$ in Eq.~(\ref{matr_eqn}) are expressed in terms of the
function $G$ as
\begin{equation}
\psi_j(t) = [q_h-q_{opt}(t_h)]G_{j2}(t,t_h)/G_{22}(t_h,t_h).
\label{explicit}
\end{equation}
\noindent
Clearly, the functions $\psi_j$ are proportional to the distance
of the point $q_h$, for which the prehistory probability density is
sought, to the optimal fluctuational path $q_{opt}(t_h|q_f,t_f)$.

Using Eq.~(\ref{matrix}), the prehistory probability density can be
expressed in terms of the functions $\psi_{1,2}$, and then in terms of
the Green function $G$,
\begin{eqnarray}
p_h(q_h,t_h|\,q_f,t_f)&=&(2\pi
D\sigma_h^2(t_h|q_f,t_f))^{-1/2}\exp\left(-{[q_h-q_{opt}(t_h|q_f,t_f)]^2\over
2D\sigma_h^2(t_h|q_f,t_f)}\right)\label{gauss}\\
&&\sigma_h^{2}(t_h|q_f,t_f) =
%S\left[-i{G_{12},\tilde\psi_2]/[q_h-q_{opt}(t_h)]^2
G_{22}(t_h,t_h).\nonumber
\end{eqnarray}
\noindent
The distribution $p_h$ is Gaussian, with a maximum on the optimal
path.  It is seen from Eqs.~(\ref{explicit}), (\ref{gauss}) that the
variance of the distribution is given immediately by the component of
the Green function $G_{22}(t_h,t_h)$. For very large $t_f-t_h$ the
variance becomes independent of $t_f$, and Eq.~(\ref{gauss}) describes
the stationary distribution about the stable state.

Alternatively, the problem of the variance of the prehistory probability
distribution can be solved in terms of the eigenfunctions and
eigenvectors of the appropriate Hamiltonian. For white-noise driven
systems this method was discussed earlier \cite{corrals}.

We note that the above analysis makes it also possible to investigate
the ``post-history'' probability distribution: the distribution of the
paths of the system {\it after} it has fluctuated to a remote state. For
white-noise driven systems, this distribution peaks at the path along
which the system, prepared initially in the state $q_f$, goes down to the
stable state $q^{(0)}$, as clearly demonstrated experimentally by
Luchinsky and McClintock \cite{nature}. As discussed below
Eq.~(\ref{f_general}),  colored noise is not turned off once it has
driven the system to a given state, and therefore it affects the motion
of the system after the state $q_f$ has been reached.

Formulation of the post-history problem requires changing in all
above equations to integration over the paths $q(t)$ and the auxiliary
field $k(t)$ for $t$ varying from $-\infty$ to $\infty$ (instead of
$-\infty$ to $t_f$), with the condition that the paths go through the
state $q_f$ at the instant $t_f$. The final answer is again given by
Eq.~(\ref{gauss}), and the distribution peaks at the most probable
path for the motion of the system after the state $q_f$ has been reached.

\section{Prehistory probability distribution close to the stable
state}

Explicit expressions for the prehistory probability density can be
obtained if the final point $q_f$ lies close enough to the stable state
$q^{(0)}(t_f)$, in which case the equations for optimal paths of the
system and the force are linear. In the prehistory problem, to the
lowest order in $|q_f-q^{(0)}(t_f)|$, one can neglect the term in
(\ref{kernel}) with $\lambda_{opt}K''\propto q_f-q^{(0)}(t_f)$ (cf.
(\ref{G(t_f)})), and can also replace $K'$ by its value for the stable
state.  Eq.~(\ref{matr_eqn}) may then be immediately integrated. After
straightforward but somewhat tedious calculations one obtains the
following expression for the reduced variance of the distribution $p_h$:
\begin{equation}
\sigma_h^2(t_h|q_f,t_f) =
\left[g(t_f,t_f)g(t_h,t_h)-g^2(t_f,t_h)\right]/g(t_f,t_f)
\label{sigma_lin}
\end{equation}
\noindent
(the function $g(t,t')$ is defined in Eq.~(\ref{G(t_f)})).

We emphasize that Eq.~(\ref{sigma_lin}) applies for an {\it arbitrary}
shape of the power spectrum of the noise (however, we assume that the
function $\bar\phi(\tau)$ decays at least exponentially for
$|\tau|\rightarrow \infty$). It applies also for an arbitrary periodic
driving. It may be further simplified in the absence of periodic
driving, in which case $K' = -\alpha =$ const, with $\alpha > 0$, and
we obtain from (\ref{G(t_f)})
\begin{equation}
g(t,t') = {1\over 2\pi}\int d\omega\,D^{-1}\Phi(\omega)(\alpha^2 +
\omega^2)^{-1}e^{i\omega(t-t')}. \label{dima}
\end{equation}
\noindent This expression makes it simple to calculate the variance of
the prehistory probability density for an arbitrary shape of the noise
power spectrum $\Phi(\omega)$. The results of these calculations for
noise with the Gaussian power spectrum centered at
zero frequency are shown in Figs.~2. It follows from this figure that
the noise color changes the broadening of the prehistory distribution
very substantially. An important consequence of Eqs.~(\ref{sigma_lin}),
(\ref{dima}) is that the variance $\sigma_h^2(t_h|q_f,t_f) \propto
(t_f-t_h)^2$ for small $t_f-t_h$.  This is qualitatively different from
the linear time dependence of $\sigma_h^2$ in the case of white noise,
known from \cite{prehistory,corrals}. The reason is that, for systems
driven by colored noise, the mean square displacement over the time $\Delta
t$, which is small compared to the noise correlation time $t_{corr}$, is
proportional to $(\Delta t)^2$, not to $\Delta t$, as for white-noise
induced diffusion.
\begin{figure}[h]
\includegraphics[angle=-90,width=3.5in]{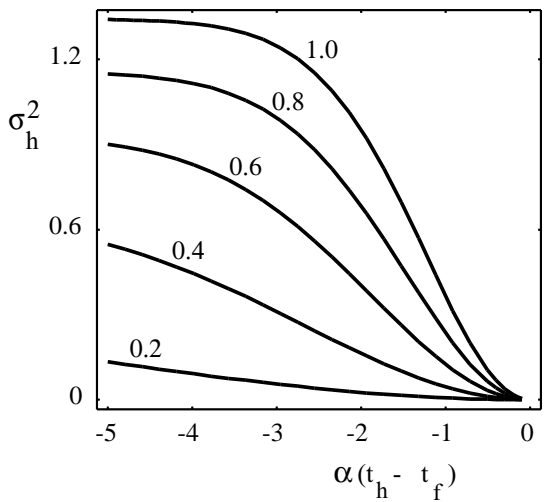}
\caption{Time dependence of the scaled variance $\sigma_h^2$
of the prehistory probability density in a linear system driven by noise
with the Gaussian power spectrum $\Phi(\omega) =
D\exp(-\omega^2/2\alpha^2\sigma^2)$, for different values of the
dimensionless variance of the noise spectrum $\sigma^2$. The time is
scaled by the decrement of the system $\alpha = -K'(q^{(0)})$.}
\end{figure}

For large $t_f-t_h$, which exceeds the relaxation time of the system
and the correlation time of the noise $t_{rel},t_{corr}$, the
prehistory probability distribution goes over into the stationary
distribution which is described by the function $R(q_h,t_h) =
[q_h-q^{(0)}(t_h)]^2/2\sigma_h^2$, where $R(q_h,t_h)$ is given by
Eq.~(\ref{harmonic}).

\section{Conclusions}

In contrast to white noise driven systems, for colored noise, after
the noise has driven the system to a remote state $q_f$, it does not
become small at a time. As the noise decays it drives the system
further along a certain path. This path differs from the path which
the system would follow if it were prepared in the state $q_f$ ``by
hand'', not as a result of the fluctuation.

In the present paper we provided a general formulation of the problem of
large fluctuations in systems driven by Gaussian noise. This formulation
makes it possible to describe optimal fluctuational paths of the system,
and also to evaluate the width of the tube of fluctuational paths that
arrive at a given target state. The latter is done using the prehistory
probability distribution for the system to have passed through a given
point on its way to the target state. The tube of the fluctuational
paths is centered at the optimal path. Evaluation of the width of the
tube has been reduced to solution of a
linear equation. Explicit results have been obtained for the
fluctuations in the linear range close to the stable state, and the
effect of noise color in this domain has been analyzed.

\noindent

\end{document}